\documentclass[aps,preprint,showpacs,superscriptaddress,groupedaddress]{revtex4}  
\usepackage{graphicx}  
\usepackage{dcolumn}   
\usepackage{bm}        
\usepackage{amssymb}   
\usepackage{amsmath}
\usepackage[colorlinks=true, citecolor=blue, urlcolor=blue, linkcolor = blue]{hyperref}
\usepackage{bm}
\usepackage{color}
\usepackage{bookmark}
\usepackage{tabularx}
\usepackage{mathtools}
\usepackage{microtype}
\usepackage{cancel}
\usepackage{relsize}
\usepackage{float}
\usepackage[caption = false]{subfig}
\usepackage{physics}

\begin{document}

%

\title{Bragg's law for X-ray scattering by quantum thermodynamic time crystals, Q-balls, as manifestation of the mechanism of  High-T$_c$ superconductivity}






\author{S. I. Mukhin}
\affiliation{Theoretical Physics and Quantum Technologies Department, NUST ``MISIS'', Moscow, Russia}
\date{\today}

\begin{abstract}
Proposed by the author Q-ball mechanism of the pseudogap state and high-Tc superconductivity in cuprates was recently supported by micro X-ray diffraction data in HgBa$_2$CuO$_{4+y}$. This provides a remarkable opportunity to investigate X-ray diffraction produced by the quantum thermodynamic time crystals, a direct embodiment of those are just the Euclidean Q-balls considered in the aforementioned theory. Simultaneously, it is also demonstrated that T-linear temperature dependence of electrical resistivity in the Q-ball phase arises due to scattering of electrons on the condensed charge/spin fluctuations inside the Q-balls. This gives a clue concerning possible mechanism of T-linear behaviour of electrical resistivity in the strange metal phase of high-T$_c$ cuprates. For this purpose the Green's functions of X-ray photons and fermions scattered by the Q-balls in the pseudogap phase of high-Tc superconductors are calculated using the Feynman diagrammatic  technique. The Bragg's peaks intensity, provided by the imaginary part of the retarded photon Green's function, is calculated. In total, obtained results describe X-ray and electron scattering on the finite size Q-ball of CDW/SDW density, that oscillates with bosonic frequency $\Omega=2\pi nT$ in Matsubara time, i.e. the quantum thermodynamic time crystal. It is found, that theoretical results are in good correspondence with X-ray diffraction data in HgBa$_2$CuO$_{4+y}$ reported recently. The T-linear dependence of electrical resistivity arises due to inverse temperature dependence of the Q-ball radius as function of temperature.
\end{abstract}


\maketitle

\section{Introduction}
A recent theory of a Q-ball mechanism of the pseudogap (PG) phase and high-T$_c$ superconductivity in cuprates \cite{Mukhin(2022), Mukhin(2022_1)}was proposed to account theoretically for the most salient features of  these and newly found compounds. As the next step the Q-balls theory predictions for X-ray scattering \cite{Mukhin(2023)} were found in favourable accord with the X-ray diffraction experimental results on high-Tc cuprate superconductors in the pseudogap phase \cite{campi22,campi}. This provides a remarkable opportunity to investigate X-ray diffraction produced by the quantum thermodynamic time crystals, a direct embodiment of those are just the Euclidean Q-balls considered in the aforementioned theory. For this purpose the X-ray 'photon' Green's function scattered by the Q-balls in the pseudogap phase of high-Tc superconductors is calculated below using the same Feynman diagrammatic  technique that was previously used for random impurity scattering in metal alloys, but with some important adjustment. In the derivation a toy model for "scalar" photonic Green's function is used for simplicity. As a result, the Bragg's peak intensity, provided by the imaginary part of the retarded 'photon' Green's function, is calculated. Hence, presented results describe X-ray scattering on the finite size Q-ball of CDW/SDW density, that oscillates with bosonic frequency $\Omega=2\pi nT$ in Matsubara time, i.e. the quantum thermodynamic time crystal. 
A few remarks are in order to simplify the job for the general reader. Namely, turning back to the Q-ball picture, we mention that an essential prerequisite for the Q-balls emergence is the attraction between condensed elementary bosonic spin-/charge-density-wave excitations. It  is self-consistently triggered by the formation of Cooper-pairs condensates inside Euclidean Q-balls~\cite{Mukhin(2022)}. Hence, the binding of the fermions into Cooper/local pairs inside the Q-balls occurs via an exchange with semiclassical density fluctuations of a finite amplitude below a high enough temperature T$^*$. The~latter is of the order of the excitation `mass', i.e.,~proportional to the inverse of  the correlation length of the short-range spin-/charge-density-wave fluctuations.  The~Q-ball charge Q counts the number of condensed elementary bosonic excitations forming the finite amplitude spin-/charge-density wave inside the Q-ball volume. The~amplitude of the Q-ball fluctuation lies in the vicinity of the local minimum of the free energy of the Q-ball, thus making it stable. Euclidean Q-balls arise due to the global invariance of the effective theory under the $U(1)$ phase rotation of the Fourier amplitudes of the spin-/charge-density fluctuations, leading to the conservation of the `Noether charge' Q in Matsubara time. This is reminiscent of the Q-balls formation in the supersymmetric standard model, where the Noether charge responsible for the baryon number conservation in real time is associated with the ${U}(1)$ symmetry of the squarks field~\cite{Rosen, Coleman (1985), Lee and Pang (1992)}. Contrary to the squark Q-balls, the~Euclidean Q-balls arise at finite temperature T$^*$ and the phase of the dominating Fourier  component of the spin-/charge-density wave fluctuation rotates with bosonic Matsubara frequency $\Omega=2\pi T$ in the Euclidean space time. Simultaneously, the~local minimum of the Q-ball potential energy located at the finite value of the modulus of the Fourier amplitude arises due to the local/Cooper pairing~\cite{Mukhin(2022_1)}. A~`bootstrap' condition 
is an exchange with fluctuations of a finite amplitude that causes the local/Cooper pairing of fermions inside Q-balls already at high temperatures. An~idea of a semiclassical `pairing glue' between fermions in cuprates, but~for an itinerant case, was proposed earlier in~\cite{Mukhin (2018)}. Hence, the  proposed  superconducting pairing mechanism inside Q-balls is distinct from the usual phonon- \cite{elis} or spin-fermion coupling models~\cite{Chubukov (2003)} considered previously for high-T$_c$ cuprates, based upon the exchange with infinitesimal  spin- and charge-density fluctuations~\cite{Seibold} or~polarons~\cite{Bianconi (1994)} in the usual Fr\"{o}hlich~picture. 

\section{Quintessence of Euclidean Q-Balls~Picture}
In order to derive the explicit relation for the Q-ball charge conservation, one may use~\cite{Mukhin(2022), Mukhin(2022_1)} a simple model Euclidean action ${S}_{M}$ with a scalar complex field $M(\tau,{\bf{r}})$, written as:
\begin{eqnarray}
S_{M}=\int_0^{\beta}\int_Vd\tau d^D{\bf{r}}\dfrac{1}{g}\left\{ |\partial_{\tau}M|^2 +s^2 |\partial_{{\bf{r}}}M|^2 + {\mu _0 ^2 }{|M|}^2+ gU_{f}(|M|^2)\right\},\; M\equiv M(\tau,{\bf{r}})\,, \label{Eu}
\end{eqnarray}
\noindent where $s$ is bare propagation velocity, and~the `mass' term $\mu_0^2\sim 1/\xi^2$ is responsible for finite correlation length $\xi$ of the fluctuations. Effective potential energy $U_{f}(|M|^2)$, as~was first derived in~\cite{Mukhin(2022), Mukhin(2022_1)}, depends on the field amplitude $|M|$ and contains charge-/spin-fermion coupling constant $g$ in front.  $M(\tau+1/T,{\bf{r}})=M(\tau,{\bf{r}})$ is periodic function of Matsubara time at finite temperature $T$ \cite{agd} and~may be considered, e.g.,~as an amplitude of the SDW/CDW fluctuation with wave vector ${\bf{Q_{DW}}}$:
\begin{eqnarray}
&&{M}_{\tau,{\bf{r}}}={M}(\tau,{\bf{r}})e^{i{\bf{Q_{DW}\cdot r}}}+{M(\tau,{\bf{r}})}^{*}e^{-i{\bf{Q_{DW}\cdot r}}},\; \nonumber\\
&&M(\tau,{\bf{r}})\equiv |M({\bf{r}})|e^{-i\Omega\tau},\;\Omega=2\pi nT,\; n=1,2,...
\label{SDWQ0}
\end{eqnarray}
\noindent {The} model (\ref{Eu}) is $U(1)$ invariant under the global phase rotation $\phi$: $M\rightarrow Me^{i\phi}$. Hence, corresponding  `Noether charge' is conserved along the Matsubara time axis. The~`Noether charge' conservation makes possible Matsubara time periodic, finite volume Q-ball semiclassical solutions, that otherwise would be banned in D$>2$ by Derrick theorem~\cite{Derrick} in the static case. Previously, Q-balls were introduced by Coleman~\cite{Coleman (1985)} for Minkowski space in QCD and~were classified as non-topological solitons~\cite{Lee and Pang (1992)}. It is straightforward to derive classical dynamics equation for the field $M(\tau,{\bf{r}})$ from Equation~(\ref{Eu}):
\begin{eqnarray}
\dfrac{\delta S_{M}}{\delta M^*(\tau,{\bf{r}})}=-\partial^2_\tau M(\tau,{\bf{r}})-s^2\sum_{\alpha={\bf{r}}}\partial^2_\alpha M(\tau,{\bf{r}})+\mu _0 ^2 {M(\tau,{\bf{r}})}+gM(\tau,{\bf{r}})\dfrac{\partial U_f}{\partial 
|M(\tau,{\bf{r}})|^2 }=0.\label{L}
\end{eqnarray}
\noindent {It} provides conservation of the `Noether charge' Q defined via space integral of the Euclidean time component  $j_{\tau}$ of the $D+1$-dimensional `current density' $\{j_{\tau},\vec{j}\}$ of the scalar field $M(\tau,{\bf{r}})$:
\begin{eqnarray}
Q=\int_V j_{\tau}d^D{\bf{r}}\; , \label{Q}
\end{eqnarray}
\noindent where the current density is defined as:
\begin{eqnarray}
j_{\alpha}=\frac{i}{2}\left\{M^{*}(\tau,{\bf{r}})\partial_\alpha M(\tau,{\bf{r}})-M(\tau,{\bf{r}})\partial_\alpha M^{*}(\tau,{\bf{r}})\right\}, \;\alpha=\tau,{\bf{r}}.\label{J}
\end{eqnarray}

\noindent {It} is straightforward to check that charge Q is conserved for the non-topological field configurations, that occupy finite volume $V$,  i.e.,~$M(\tau,{\bf{r}}\notin V)\equiv 0$:
\begin{eqnarray}
\dfrac{\partial Q}{\partial\tau}=\dfrac{\partial}{\partial\tau}\int_V j_{\tau}d^D{\bf{r}}=-s^2\oint_{S(V)}\vec{j}\cdot d\vec{S}=0\;,\label{NN}
\end{eqnarray}
\noindent {Now}, approximating the `Q-ball' field configuration with a step function $\Theta ({\bf{r}})$:
\begin{align}
M(\tau,{\bf{r}})=e^{-i\Omega\tau}M\Theta\left\{{\bf{r}}\right\}\;;\quad \Theta({\bf{r}})\equiv\begin{cases}1 ;\;{\bf{r}}\in V ;\\
0;\;{\bf{r}}\notin V.\end{cases}
\label{step}
\end{align}
\noindent 
one finds expression for the conserved charge Q:
\begin{eqnarray}
Q=\int_V j_{\tau}d^D{\bf{r}}= \Omega M^2V. \label{Qs}
\end{eqnarray}
\noindent {This} relation leads to inverse proportionality between volume $V$ and fluctuation scattering intensity $\sim$$M^2$ of, e.g.,~X-ray radiation by the density wave inside a Q-ball, see Equation~(\ref{MV}) below. 

{It} is important to mention here that the non-zero charge Q in Equation~(\ref{Qs}) follows as a result of broken `charge neutrality' in the choice for the  SDW/CDW fluctuation in Equation~(\ref{SDWQ0}), where periodic dependence on Matsubara time $\tau$ enters via an exponential factor with a single sign frequency $\Omega$, rather than in the form of a real function, e.g.,  $\propto \cos(\Omega\tau+\phi)$. Now, in~the step-function approximation of Equation~(\ref{step}), the action $S_M$ equals:
\begin{eqnarray}
 S_{M}=\frac{1}{gT }\left\{\dfrac{Q^2}{VM^2}+ V[\mu _0 ^2M^2 + gU_{f}]\right\}, \label{SMQ}
\end{eqnarray}
\noindent where Equation~(\ref{SMQ}) is obtained using charge conservation condition Equation~(\ref{Qs}). It is remarkable that as it follows from the above expression in Equation~(\ref{SMQ}), the~Q-ball volume enters in denominator in the $\propto Q^2/V$ term. Hence, provided the $\propto V$ term is positive, there is a minimum of action $S_M$ (free energy) at finite volume $V_Q$ of a Q-ball.  Hence, volume $V_Q$ that minimises $S_{M}$ and energy $E_Q$ equal:
\begin{eqnarray}
 V_Q= \dfrac{Q}{M\sqrt{\mu _0 ^2 M^2+ gU_{f}(M)}}\;; \label{VQ}\\
 E_Q=TS^{min}_M=\dfrac{2Q\sqrt{\mu _0 ^2 M^2+ gU_{f}(M)}}{gM}=\dfrac{2Q\Omega}{g}, \label{aQ}
\end{eqnarray}
\noindent where the last equality in Equation~(\ref{aQ}) follows directly after substitution of  expression $V_Q$ from Equation~(\ref{VQ}) into Equation~(\ref{Qs}), which then expresses $V_Q$ via $Q$ and $\Omega$. As~a result, charge $Q$ cancels in Equation~(\ref{aQ}), and~the following self-consistency equation follows~\cite{Mukhin(2022_1)}:
\begin{eqnarray}
 0=(\mu _0 ^2 -\Omega^2)M^2+ gU_{f}(M). \label{self}
\end{eqnarray}
\noindent
{Another} self-consistency equation arises from solution of the Eliashberg-like equations with the SDW/CDW fluctuation field ${M}_{\tau,{\bf{r}}}$ from Equation~(\ref{SDWQ0}) playing role of the pairing boson~\cite{Mukhin(2022), Mukhin(2022_1)}. Namely, it was also demonstrated in~\cite{Mukhin(2022), Mukhin(2022_1)} that a fermionic spectral gap $g_0$ inside Euclidean Q-balls arises in the vicinity of the `nested' regions of the bare Fermi surface (corresponding to the antinodal points of the cuprates Fermi surface) and scales with the local superconducting density $n_s$ inside the Q-balls:
\begin{eqnarray}
 g_0=\sqrt{2M(M-\Omega)}\,;\; n_s =2|\Psi|^2 \approx \frac{\nu\varepsilon_0}{2}\tanh^2\frac{g_0}{2T}\tanh\frac{2 g_0}{3\varepsilon_0} ,
 \label{ns1}
\end{eqnarray}
\noindent where $|\Psi|^2$ is local/Cooper-pairs density inside Q-ball~\cite{Mukhin(2022)}, and~$\nu\varepsilon_0$ is the density of fermionic states involved in `nesting'. Substitution of  Equation~(\ref{ns1}) into expression for the Q-ball free energy drop due to pairing of fermions leads to the following expression for the pairing-induced effective potential energy $U_{eff}(M)$ of SDW/CDW field~\cite{Mukhin(2022), Mukhin(2022_1)}:
\begin{align}
&U_{eff}(M)\equiv\mu _0 ^2 M^2+gU_f=\mu _0 ^2 M^2 -\dfrac{4g\nu \varepsilon_0\Omega}{3}I\left(\dfrac{M}{\Omega}\right)\;,\;M\equiv |M(\tau)|\label{UFO1}\\
&I\left(\dfrac{M}{\Omega}\right)= \int_1^{M/\Omega}d\alpha\dfrac{\alpha \sqrt{2\alpha \left(\alpha -1\right)}}{(1+8\alpha \left(\alpha -1\right))}\tanh{\dfrac{\sqrt{2\alpha\left(\alpha-1\right)}\Omega}{\varepsilon_0}}\tanh{\dfrac{\sqrt{2\alpha\left(\alpha-1\right)}\Omega}{2T}}.
\label{UFOI}
\end{align}

\begin{figure}[H]
\includegraphics[width=0.45\linewidth]{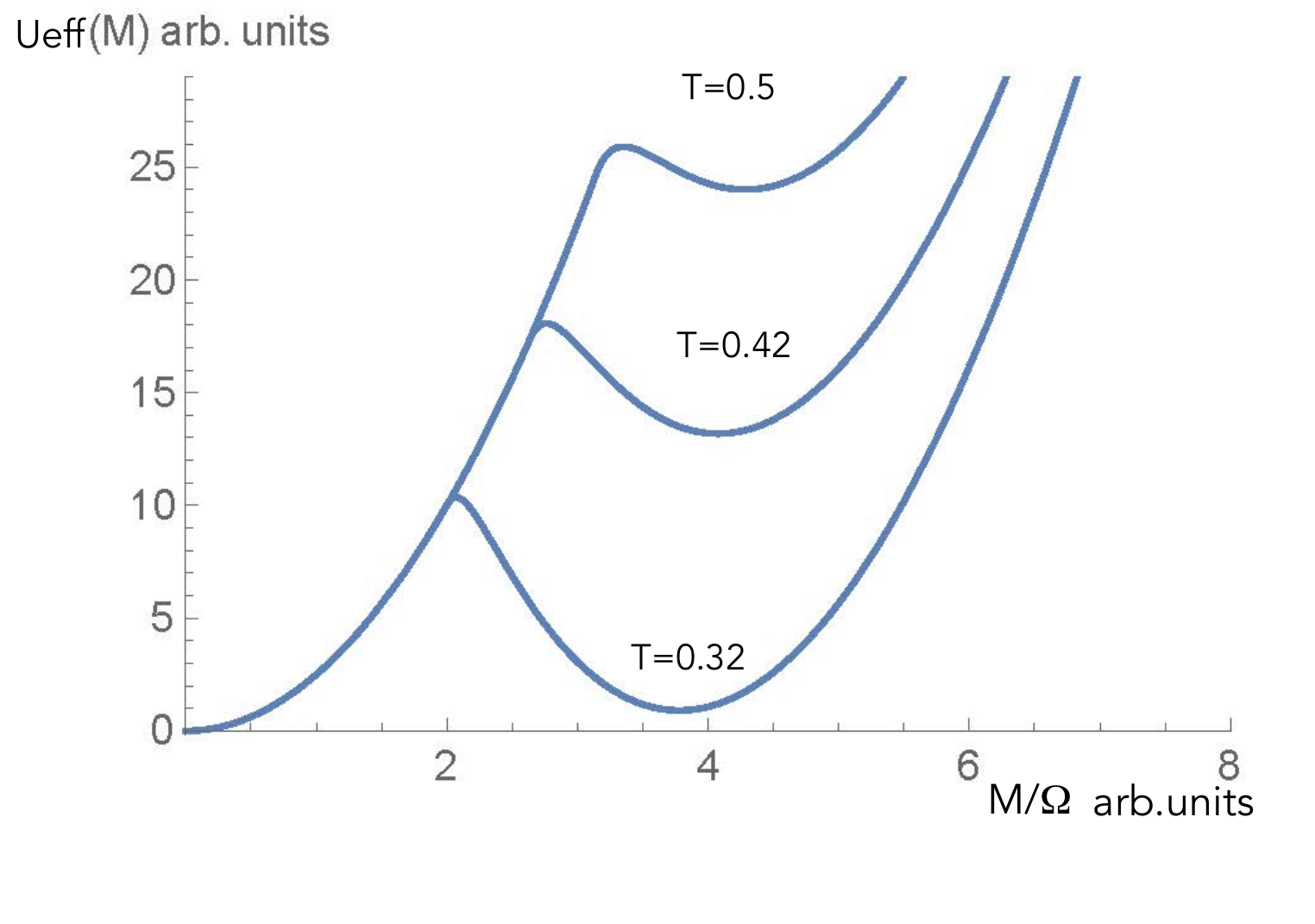}\vspace{0.1cm}
\caption{The plots of $U_{eff}(M)$ at different normalised temperatures $T/T^*$ manifesting characteristic Q-ball local energy minimum at finite amplitude due to condensation of local/Cooper pairs inside Q-balls, obtained from Equations~(\ref{UFO1}) and  (\ref{UFOI}), see text.\label{triple}}
\end{figure}

  The plot of $U_{eff}(M)$ vs. $M/\Omega$ for different temperatures $T \leq T^*=\mu_0/2\pi$ is presented in Figure  \ref {triple}.
The figure manifests characteristic Q-ball local minimum at finite amplitude that, in contradistinction with the squarks theory~\cite{Coleman (1985)}, is produced here by condensation of superconducting local/Cooper pairs inside the CDW Q-balls, first arising at temperature $ T^*$. The~minimum deepens down when temperature decreases to $T=T_c$, at~which Q-ball volume becomes infinite and bulk superconductivity sets~in.  
\section{Summary of Theoretical Predictions for~Q-Balls}
Summarising,   Equation~(\ref{Eu}) was used to describe effective theory of  the Fourier components of  the leading Q-ball (i.e., short-range) SDW/CDW fluctuations. Explicit expression for $U_{f}(|M(\tau,{\bf{r}})|)$ was derived and investigated in detail previously~\cite{Mukhin(2022), Mukhin(2022_1)}  by integrating out Cooper/local-pairs fluctuations in the `nested' Hubbard model with charge-/spin-fermion interactions. As~a result, Q-ball self-consistency Equation~(\ref{self}) was solved and investigated, and~it was established that  Euclidean Q-balls  describe stable semiclassical short-range charge/spin-ordering fluctuations of finite energy that appear at finite temperatures below some temperature T$^*$, found to be $T^*\approx \mu_0/2\pi$ \cite{Mukhin(2022), Mukhin(2022_1)}. Next, it was also found that transition into pseudogap phase at the temperature T$^*$ is of 1st order with respect to the amplitude $M$ of the Q-ball  SDW/CDW fluctuation and of 2nd order with respect to the superconducting gap $g_0$.  In~particular, the~following temperature dependences of these characteristics of the Q-balls were derived from Equations~(\ref{self}), (\ref{ns1}), and (\ref{UFOI}) in the vicinity of  the transition temperature T$^*$ into pseudogap phase~\cite{Mukhin(2022_1)} for the CDW/SDW amplitude:
\begin{eqnarray}
M=\Omega\left(1+ \left( \dfrac{T^*-T}{\mu_0}\right)^{\frac{2}{5}}\left(\dfrac{15\mu^2 _0}{4\sqrt{2}g\nu}\right)^{\frac{2}{5}}\right),\quad T^*=\dfrac{\mu_0}{2\pi}\,,
 \label{Mstar}
\end{eqnarray}
 \noindent  and for the pseudogap $g_0$:
\begin{equation}
g_0^2= \left( {T}^*-T\right)^{\frac{2}{5}}{\Omega}^2\left(\dfrac{15\mu _0}{g\nu}\right)^{\frac{2}{5}},
\label{g0star}
\end{equation}
\noindent which follows after substitution of Equation~(\ref{Mstar}) into Equation~(\ref{ns1}). These dependences are plotted in Figure~5b in~\cite{Mukhin(2022_1)}.
\subsection{Temperature dependences of Q-ball parameters close to $T^*$}
Strikingly, but it follows from Equation (\ref{g0star}),  that micro X-ray diffraction data also allow to infer an emergence of superconducting condensates inside the Q-balls below T$^*$. The~reason is in the inflation of the volume, which is necessary to stabilise the superconducting condensate at vanishing density. Indeed, this is the most straightforward to infer from linearised Ginzburg--Landau (GL) equation~\cite{aaa} for the superconducting order parameter $\Psi$ of a Q-ball of radius $R$ in the spherical coordinates:
\begin{eqnarray}
-\dfrac{\hbar^2}{4m}\ddot{\chi}={bg_0^2}\chi\;; \quad\Psi(\rho)=\dfrac{C\chi(\rho)}{\rho}\;;\quad \Psi(R)=0, \label{GL}
\end{eqnarray} 
\noindent where $g_0^2$ from Equation~(\ref{ns1}) substitutes GL parameter $a=\alpha \cdot(T_c-T)/T_c$ modulo dimensionful constant $b$ of GL free energy functional~\cite{aaa}. Then, it follows directly from solution of Equation~(\ref{GL}):
\begin{eqnarray}
{\chi}\propto \sin(k_n\rho)\;;\quad Rk_n=\pi n,\;;\quad n=1,2,...,
\label{chik}
\end{eqnarray}
\noindent that Equation~(\ref{GL}) would possess solution (\ref{chik}) with the eigenvalue ${bg_0^2}$ only if  the Q-ball radius is greater than $R_{min}$:
\begin{eqnarray}
&\dfrac{\hbar^2}{4m}\left(\dfrac{\pi}{R_{min}}\right)^2= {bg_0^2}. \label{Rm}
\end{eqnarray}
\noindent
Hence, due to conservation condition Equation~(\ref{Qs}), charge Q  should obey the following condition:
\begin{eqnarray}{Q}\geq Q_{min}\equiv{\Omega M^2}(R_m)^3= {\Omega M^2}\dfrac{(\pi\hbar)^3}{g_0^3(4mb)^{3/2}}\;. 
\label{Qm}
\end{eqnarray}
\noindent
This would have an immediate influence on the temperature dependence of the most probable value of charge Q. The~letter value could be evaluated using expression for the Q-ball energy Equation~(\ref{aQ}): $E_Q={2Q\Omega}/{g}$ obtained in~\cite{Mukhin(2022)}. Then, Boltzmann distribution of energies of the Q-balls `gas' indicates that the most numerous, i.e.,~the most probable to occur, Q-balls are those with the smallest possible charge Q, and their respective population (overage) number $\bar{n}_{Q}$ in unit volume of the sample is:
\begin{eqnarray}
 \bar{n}_{Q}\propto\dfrac{1}{V}G_{Q}\exp{\left\{-\dfrac{E_{Q}}{k_BT}\right\}}=G_{Q}\exp{\left\{-\dfrac{2Q\Omega}{gk_BT}\right\}} \;,
 G_{Q}=\dfrac{V}{V_{Q}},
 \label{LL5}
 \end{eqnarray}  
 \noindent
 where $G_{Q}$ counts the number of  possible Q-ball `positions' in the sample of volume $V$, $V_{Q}$ being Q-ball volume, and~$\Omega/k_BT=2\pi$. Hence, Equation~(\ref{LL5}) indicates that the Boltzmann's exponent is greater for smaller $Q$. On the other hand,  due to accommodated superconducting condensates inside the Q-balls, their Noether charge Q is limited from below by Q$_{min}$, as demands Equation (\ref{Qm}). Substituting into Equations (\ref{Rm}) and (\ref{Qm}) temperature dependences of $M$ and $g_0$ from Equations (\ref{Mstar}) and (\ref{g0star}), one finds:
\begin{eqnarray}
&R_{min}=\dfrac{1}{\Omega (T^*-T)^{1/5}}\dfrac{\pi\hbar}{\sqrt{4mb}}\left(\dfrac{g\nu}{15\mu_0}\right)^{1/5}\,;\label{RmT}\\
&M^2=\Omega^2\left(1+ \left( \dfrac{T^*-T}{\mu_0}\right)^{\frac{2}{5}}\left(\dfrac{15\mu^2 _0}{4\sqrt{2}g\nu}\right)^{\frac{2}{5}}\right)^2\,;
\label{AmT} \\
&Q_{min}=\left(1+ \left( \dfrac{T^*-T}{\mu_0}\right)^{\frac{2}{5}}\left(\dfrac{15\mu^2 _0}{4\sqrt{2}g\nu}\right)^{\frac{2}{5}}\right)^2 \dfrac{(\pi\hbar)^3}{(4mb)^{3/2}(T^*-T)^{3/5}}\left(\dfrac{g\nu}{15\mu_0}\right)^{3/5}
 \label{QmT}
\end{eqnarray}

\noindent
An immediate measurable consequence of the Q-ball charge conservation in the form of Eq.~(\ref{Qs}) would be inverse correlation between Q-ball volume $V_Q=4\pi R_Q^3/3$ and CDW/SDW amplitude squared $M^2$ at fixed temperature $T=\Omega/2\pi$. This anticorrelation might be extracted e.g. from experimental X-ray scattering data \cite{campi22} in the form of dependence of the amplitude $A\sim M^2$ of X-ray scattering peak on its width in momentum space $\Delta k \sim 1/R_Q\sim V_Q^{-1/3}$ in the pseudogap phase of high-T$_c$ cuprates\cite{Mukhin(2022)}. In order to make a precise  prediction one has to derive X-ray scattering cross-section by Q-balls.
    Taking into account exponential dependence of the Boltzmann distribution of the energies of the Q-balls on their `Noether charge' Q and their respective population (overage) number $\bar{n}_{Q}$ in Eq. (\ref{LL5}), one may fix $Q=Q_{min}$ close enough to the transition temperature $T^*$ in the farther derivations of the X-ray scattering cross-section by the Q-balls presented below.  
     
\section{Bragg's law for X-ray scattering by Q-balls} 
\subsection{Photon Green's function in Q-balls gas}
\begin{figure}[H]
\includegraphics[width=0.45\linewidth]{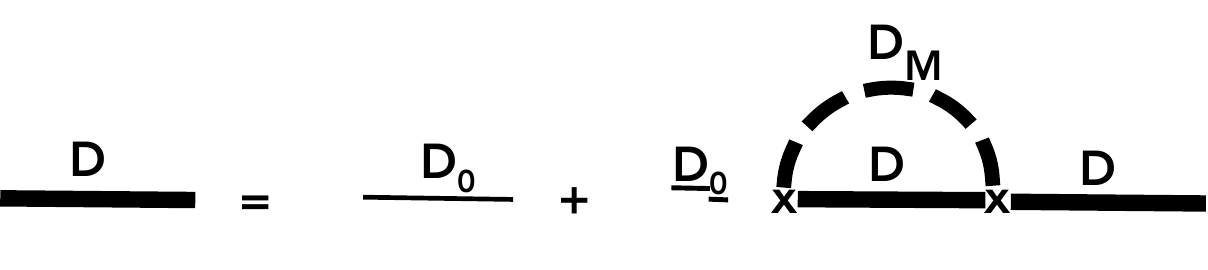}\hspace{0.2cm}
\caption{ The Dyson equation for X-ray scattering by Q-balls of CDW/SDW bosonic field : the dashed line is CDW/SDW Q-ball bosonic Euclidean field correlator $D_M^Q$ averaged over coordinates of Q-ball's centres in a crystal and Matsubara time zero-origin $\tau_0$. Heavy and thin lines are photon's temperature Green's functions  $D({r}-{r}')$ and $D_0({r}-{r}')$ respectively. Crosses are vertices of X-ray photon-Q-ball field interaction $\gamma$.} 
\label{Dys}
\end{figure}

Using phonon-like ansatz for the photon Green's function $D_0(\tau,{\bf{r}}) $ in ideal crystal and then introducing photon scattering by the Q-ball scalar field $M(\tau,{\bf{r}})$ defined in Eqs. (\ref{SDWQ0}), (\ref{step}), one finds the following equation after averaging over positions of the Q-ball centres in space and over zero-origin $\tau_0$ of Matsubara time, compare with random space impurity scattering technique \cite{agd}:

\begin{eqnarray}
&D({r}-{r}')= D_0({r}-{r}')+\nonumber\\
&+\gamma^2\displaystyle\sum_{Q}\bar{n}_{Q}\int d^4r_1d^4r_2D_0({r}-{r}_1)D_M^Q({r}_1-{r}_2)D({r}_1-{r}_2)D({r}_2-{r}')\,
\label{DS}\\
 &r\equiv\{\tau,{\bf{r}}\},\; \int d^4r\equiv\int_0^\beta d\tau\int d^3{\bf{r}}\,,\nonumber
\end{eqnarray}

\noindent where $\bar{n}_{Q}$ is given in Eq. (\ref{LL5}), interaction constant $\gamma$ sets the scale of the scattering amplitude of X-ray photon on the Q-ball field, and Boltzmann distribution of the Q-balls energies $E_{Q}$, given in  Eq. (\ref{aQ}), regulates via $\bar{n}_{Q}$ a density of Q-balls gas. The most probable to occur Q-balls are those with the smallest possible charge $Q=Q_{min}$, and $G_{Q}$ counts the number of  possible `positions' for a Q-ball in the sample of volume $V$, $V_{Q}$ being Q-ball's volume, and $\Omega/k_BT=2\pi$.  A Q-ball field correlator $D_M^Q(\tau,{\bf{r}})$ is obtained by using Eqs. (\ref{SDWQ0}), (\ref{step}) and averaging over origin $\tau_0$ of Matsubara time:
 
\begin{eqnarray}
D_M^Q(\tau,{\bf{r}})=2M^2\exp\{-{r}{\kappa}\}\cos{({\bf{Q}}\cdot{\bf{r}}-\Omega\tau)}\label{DM};\quad \kappa=\dfrac{1}{R_Q},\; R_Q=\left(3V_Q/4\pi\right)^{1/3} .
\end{eqnarray}

\noindent for a Q-ball of radius $R_Q$. The upper index $Q$ in $D_M^Q$ signifies dependence of parameter $\kappa$ on the charge $Q$ in accord with Eqs. (\ref{Qs}), (\ref{VQ}).  Equation (\ref{DS}) in diagrammatic form is presented in Fig. \ref{Dys}.

\noindent In order to pass to momentum representation in Eq. (\ref{DS}) one has to calculate the Fourier transform of the Q-ball field correlator $D_M^Q(\tau,{\bf{r}})$ defined in Eq. (\ref{DM}):

\begin{eqnarray}
&D_M^Q(\omega, {\bf{q}})= {2M^2}{\int^{1/T}_0} d\tau\int d^3{\bf{r}}\cos{({\bf{Q}}\cdot{\bf{r}}-\Omega\tau)}\exp\{-r\kappa\ -i({\bf{q}}\cdot{\bf{r}}-i\omega\tau) \}=\nonumber \\
&=\dfrac{8\pi M^2\kappa}{T}\left[\dfrac{\delta_{\omega,\Omega}}{(({\bf{Q}}-{\bf{q}})^2+\kappa^2)^2}+\dfrac{\delta_{\omega,-\Omega}}{(({\bf{Q}}+{\bf{q}})^2+\kappa^2)^2}\right]
\label{DDQ}
\end{eqnarray}

\noindent where  $\delta_{i,j}$ is Kronecker delta. Then, Eq.(\ref{DS}) takes the following form in momentum representation:

\begin{eqnarray}
&D^{-1}(\omega, {\bf{p}})= {D_0}^{-1}(\omega, {\bf{p}})-\gamma^2T\displaystyle V\sum_{\omega',Q}\bar{n}_{Q}\int d^3{\bf{q}}D_M^Q(\omega', {\bf{q}}) D(\omega-\omega', {\bf{p}}-{\bf{q}})\,;\label{DDp}\\
 &{D_0}^{-1}(\omega, {\bf{p}})= -\dfrac{\omega^2+c^2p^2}{c^2p^2}\,,\;\omega=2\pi nT\,;n=\pm 1,\pm2,...\,.
\label{DDp0}
\end{eqnarray}
\noindent Here $c$ is light (X-ray radiation) velocity and linear dispersion $\omega_0(p)=cp$ is assumed for simplicity. Hence, after tedious, but straightforward calculation one finds photon's self-energy to the second order in coupling strength $\gamma$  :

\begin{eqnarray}
&D^{-1}(\omega, {\bf{p}}) = {D_0}^{-1}(\omega, {\bf{p}})+16\pi^2\gamma_{Q}^2M^2I_2(\omega)\,;\quad I_2(\omega)=I_{+}+I_{-}\,;
\label{DDp2}\\
&I_{+}=\dfrac{\kappa\dfrac{|\omega-\Omega|^3}{c^3}+\frac{1}{4}\left[\left(\dfrac{\omega-\Omega}{c}\right)^2(\Delta^2-3\kappa^2)+(\Delta^2+\kappa^2)^2\right]}{\left(\Delta^2+\kappa^2-\left(\dfrac{\omega-\Omega}{c}\right)^2\right)^2+4\left(\dfrac{\omega-\Omega}{c}\right)^2\Delta^2}\,;\quad \Delta\equiv|{\bf{Q}}-{\bf{p}}|\,; \label{I2}\\
& I_{-}=I_{+}(-\Omega,-{\bf{Q}}), \label{Qp}\\
&\gamma_{Q}^2\equiv V\bar{n}_{Q}\gamma^2\propto\gamma^2VV_{Q_{min}}^{-1}\exp{\left\{-\dfrac{2Q_{min}\Omega}{gk_BT}\right\}},
\end{eqnarray}
\noindent where summation over Q-ball charge $Q$ in Eq. (\ref{DDp}) is approximately substituted with the number of  Q-balls of minimal charge $Q_{min}$ calculated above in Eqs. (\ref{Qm}), (\ref{QmT}).  Next, one has to continue analytically the Green's function $D$ in Eq. (\ref{DDp2}) from imaginary axis points $i\omega=2\pi nT$ to real axis $i\omega\rightarrow z$, considering for definiteness e.g. $\Omega=2\pi T>0$, in accord with Eq. (\ref{SDWQ0}).  Then expressions for the integrals $I_{\pm}$ in Eq. (\ref{Qp}) take the form:

\begin{eqnarray}
&D^{-1}(z, {\bf{p}}) =  \dfrac{z^2-c^2p^2}{c^2p^2}+16\pi^2\gamma_Q^2M^2(I_{+}(z)+I_{-}(z))\,;\label{DDp2z}\label{DZ}\\
&I_{\pm}(z)=\displaystyle \dfrac{i\kappa\dfrac{(i\Omega\mp z)^3}{c^3}-\frac{1}{4}\left[\left(\dfrac{z\mp i\Omega}{c}\right)^2(\Delta_{\pm}^2-3\kappa^2)-(\Delta_{\pm}^2+\kappa^2)^2\right]}{\left[\kappa^2+\left(\Delta_{\pm}-\left(\dfrac{z\mp i\Omega}{c}\right)\right)^2
\right]\left[\kappa^2+\left(\Delta_{\pm}+\left(\dfrac{z\mp i\Omega}{c}\right)\right)^2\right]}\;,
\label{Qpz}\\
&\Delta_{\pm}=|{\bf{Q}}\mp{\bf{p}}|.\label{DEL}
\end{eqnarray}

\begin{figure}[H]
\includegraphics[width=0.45\linewidth]{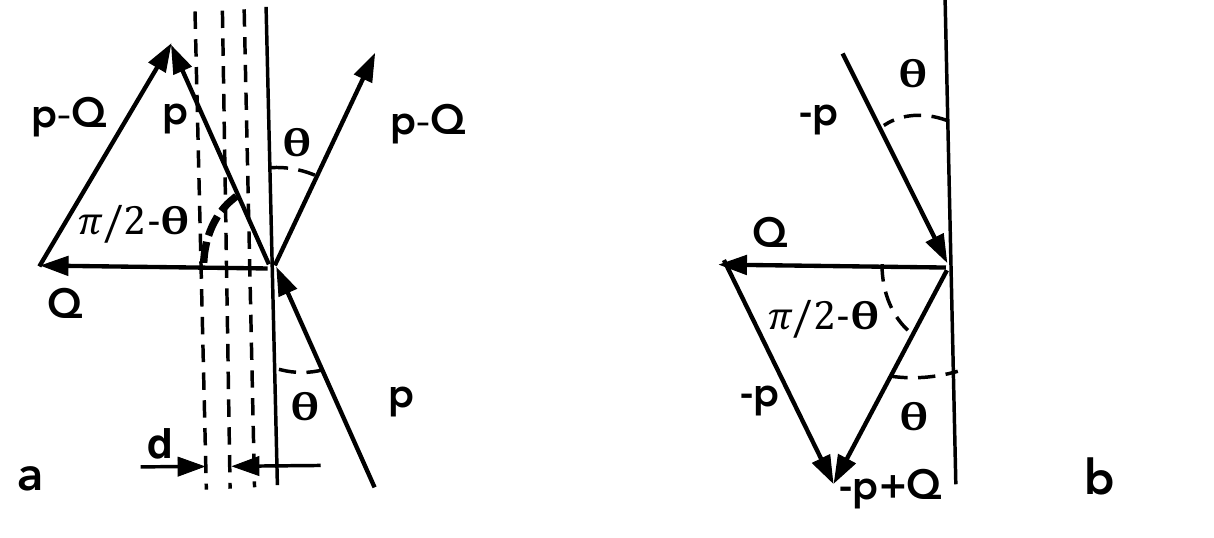}\hspace{0.2cm}
\caption{a. Bragg's reflection of incident X-ray (wave vector ${\bf{p}}$) by the Q-ball with wave-vector ${\bf{Q}}$ described with the term $I_{+}$ in Eqs. (\ref{DZ}) - (\ref{DDp2z}), see text. b. The space-time  (PT) symmetric scattering event described by term with $I_{-}$ in Eq. (\ref{DZ}).} 
\label{BGG}
\end{figure}

\noindent It is easy to check, using Eq. (\ref{DM}) for the Q-ball field correlator, that contributions due to terms $I_{\pm}$ in Eq. (\ref{Qpz}), (\ref{DEL}) are space-time  (PT) symmetric, see also Fig. (\ref{BGG}a,b). First, consider for definiteness pole of $D(z, {\bf{p}})$ when contribution due to $I_{+}(z)$ dominates and $I_{-}(z)$ is negligible. Then, remarkably, the term with $I_{+}(z)$ leads to a famous Bragg's reflection law \cite{Bragg} in the scattering configuration in Fig. (\ref{BGG}a) with finite life-time, that gives X-ray scattering intensity, see Eq. (\ref{DDp2z}) below.

\noindent  Assume that relation between frequency and momentum of the photon is weakly disturbed by scattering, i.e. $z^2\approx c^2p^2$ according to Eq. (\ref{DZ}). Then,  momentum of the incident photon minimises denominator of $I_{+}(z)$ when $|{\bf{Q}}-{\bf{p}}|-|{\bf{p}}|=0$. The latter relation, according to Fig. (\ref{BGG}) and notations therein, turns into Bragg's rule for X-rays reflection by the Q-ball:

\begin{eqnarray}
&{2}\dfrac{2\pi}{Q}\sin{\Theta}\equiv{2}d\sin{\Theta} ={\lambda_p}\equiv\dfrac{2\pi}{p}\label{BQ},
\label{Bragg1}
\end{eqnarray}
\noindent where $d=2\pi/Q$ is period of CDW/SDW, see Fig. (\ref{BGG}), and $\lambda_p$ is wave length of the X-ray photon with wave vector $p$ (Plank's constant $\hbar=1$). Expanding $I_{+}$ in Eq. (\ref{Qpz}) in the vicinity of  wave vector ${\bf{p}}$ obeying the Bragg's relation $|{\bf{Q}}-{\bf{p}}|-|{\bf{p}}|=0$ in (\ref{Bragg1}) one finds:

\begin{eqnarray}
{D_R}^{-1}(\omega, {\bf{p}})=  \dfrac{\omega^2-c^2p^2}{c^2p^2}+\dfrac{2\pi^2\gamma_Q^2M^2\left[|{\bf{Q}}-{\bf{p}}|\left(|{\bf{Q}}-{\bf{p}}|-\dfrac{\omega}{c}\right)+i\dfrac{\Omega \omega}{c^2}\right]}{\kappa^2+\left(|{\bf{Q}}-{\bf{p}}|-\dfrac{\omega}{c}\right)^2}, \label{DDp2z}
\label{ineq}
\end{eqnarray}
\noindent where $\Gamma$ is scattering intensity of X-ray photon with momentum ${\bf{p}}$ by a single Q-ball with wave-vector ${\bf{Q}}$, and real frequency variable $z$ is substituted with a common notation $\omega$ giving X-ray energy $\hbar \omega$.
The overall line shape of the X-ray scattering peak, characterised by $-\frac{1}{\pi}Im{D_R}(z, {\bf{p}})$ then reads:

\begin{eqnarray}
&-\dfrac{1}{\pi}Im{D_R}(\omega, {\bf{p}})= 
\dfrac{\Gamma(\omega, {\bf{p}})}{  \left[\dfrac{\left(\dfrac{\omega}{c}\right)^2-p^2}{p^2}+\dfrac{2\pi^2\gamma_Q^2M^2p|{\bf{Q}}-{\bf{p}}|\left(|{\bf{Q}}-{\bf{p}}|-\dfrac{\omega}{c}\right)}{\kappa^2+\left(|{\bf{Q}}-{\bf{p}}|-\dfrac{\omega}{c}\right)^2}\right]^2+\Gamma(\omega, {\bf{p}})^2 }\label{imag}\\
&\Gamma(\omega, {\bf{p}})=\dfrac{2\pi^2\gamma_Q^2M^2{\Omega \omega}}{{c^2}\left(\kappa^2+\left(|{\bf{Q}}-{\bf{p}}|-\dfrac{\omega}{c}\right)^2\right)},\label{GDp2z}
\label{QpP}\\
&\kappa^2\geq \left(|{\bf{Q}}-{\bf{p}}|-p\right)^2\,,
\end{eqnarray}
\noindent  

\subsection{X-ray diffraction pattern in Q-balls gas}

\begin{figure}[H]
\includegraphics[width=0.45\linewidth]{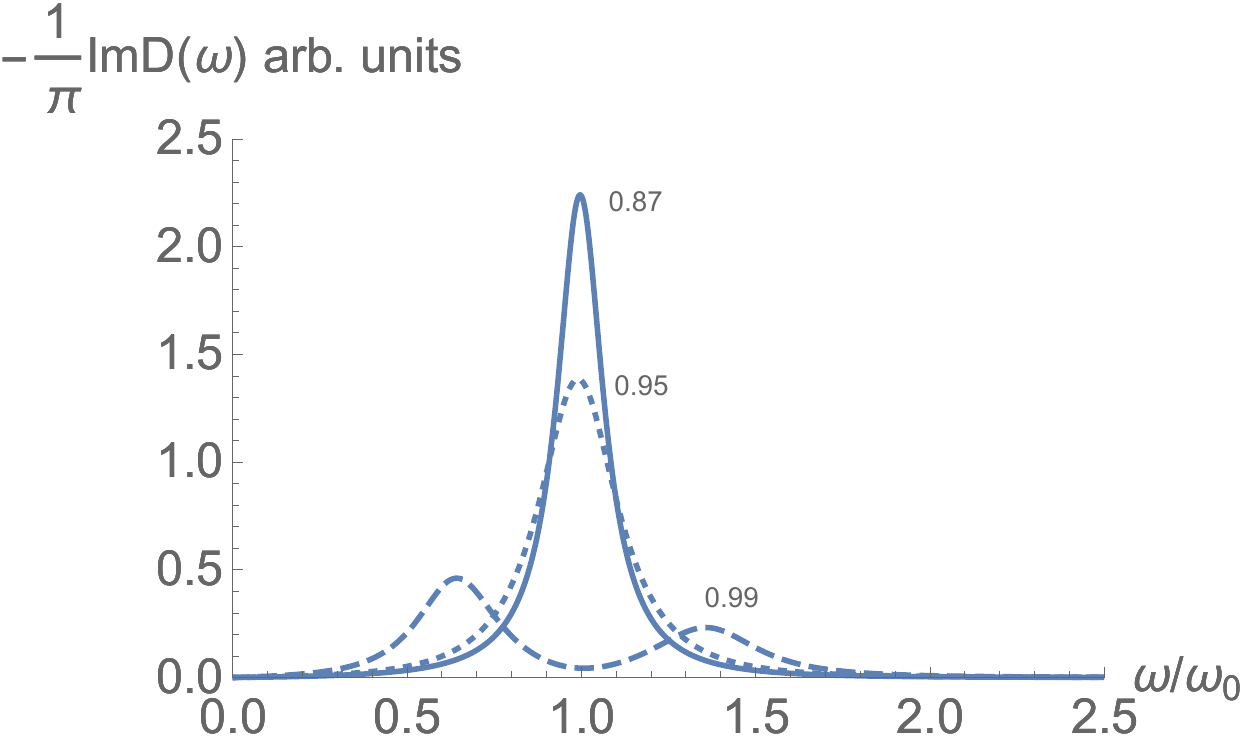}\hspace{0.2cm}
\caption{ Q-ball scattered X-ray Green's function $-\dfrac{1}{\pi}Im{D_R}(\omega, {\bf{p}})$ in Eq. (\ref{imag}) at different temperatures indicated by $T/T^*$ ratios as function of wave-vector $p$ expressed via frequency $cp=\omega$ in units of  incident X-ray frequency $\omega_0$, with dimensionless scattering amplitude $g=0.115$ defined as: $2\pi^2\gamma_Q^2M^2\Omega/T^*\equiv g(T/T^*)^3$.\label{21}}
\end{figure}

The above expression for the Green's function of the X-ray photon scattered by Q-balls is remarkable: besides information on the X-ray scattering intensity it contains the famous Bragg's reflection law \cite{Bragg}, when the Q-ball radius $R$ is big enough with respect to photon wave length $\lambda\sim 1/p$, i.e. $1/R\sim \kappa\ll p$ in Eq. (\ref{Bragg1}).  The X-ray scattering intensity is characterized by imaginary part of the Green's function $-\dfrac{1}{\pi}Im{D_R}(\omega, {\bf{p}})$ in Eq. (\ref{imag}) which provides the density of states of the scattered X-rays with energy $\hbar\omega$, see Fig. \ref{21}.

\begin{figure}
\subfloat[a) ]{\includegraphics[width = 0.49\textwidth]{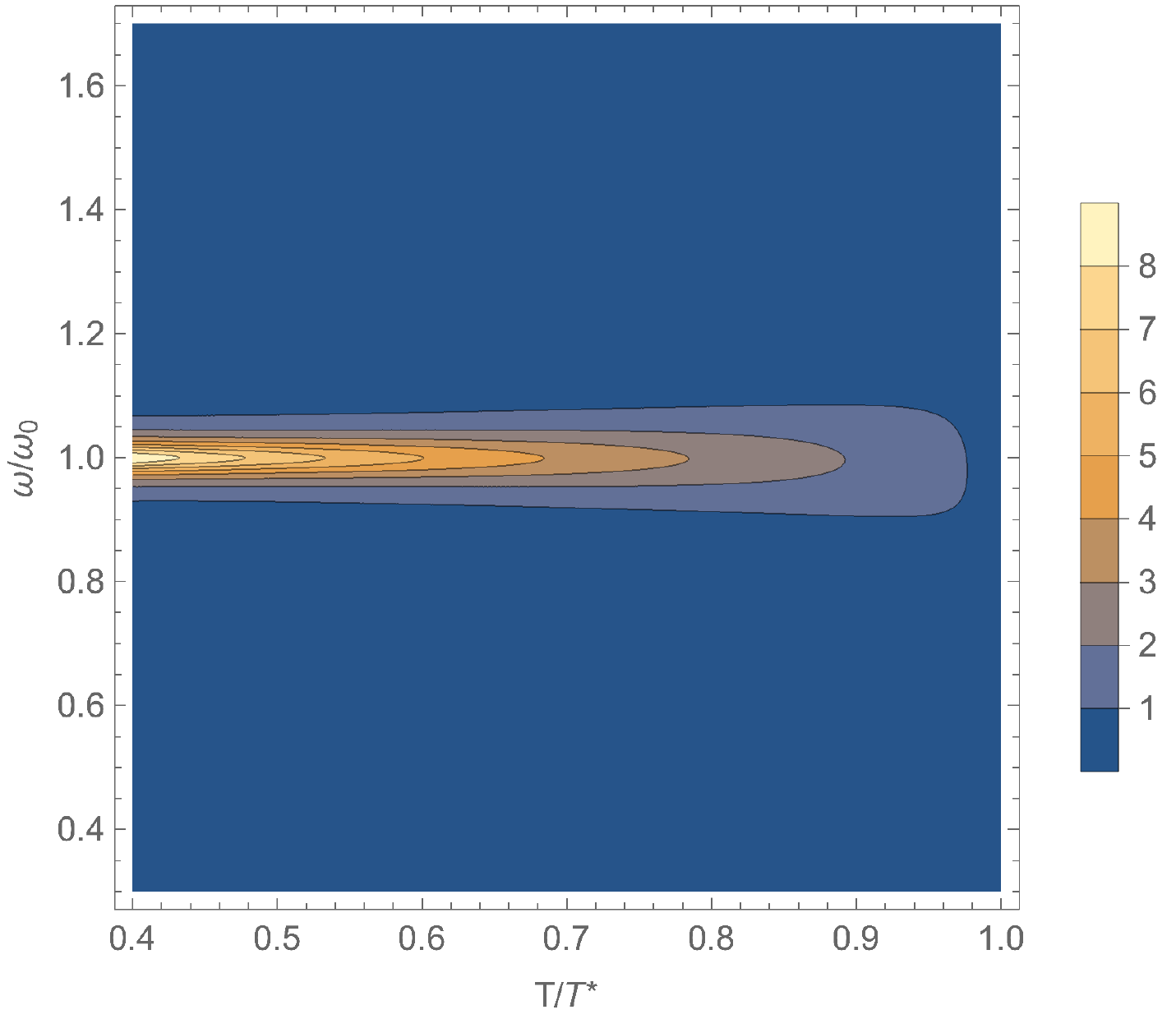}}
\subfloat[b) ]{\includegraphics[width = 0.49\textwidth]{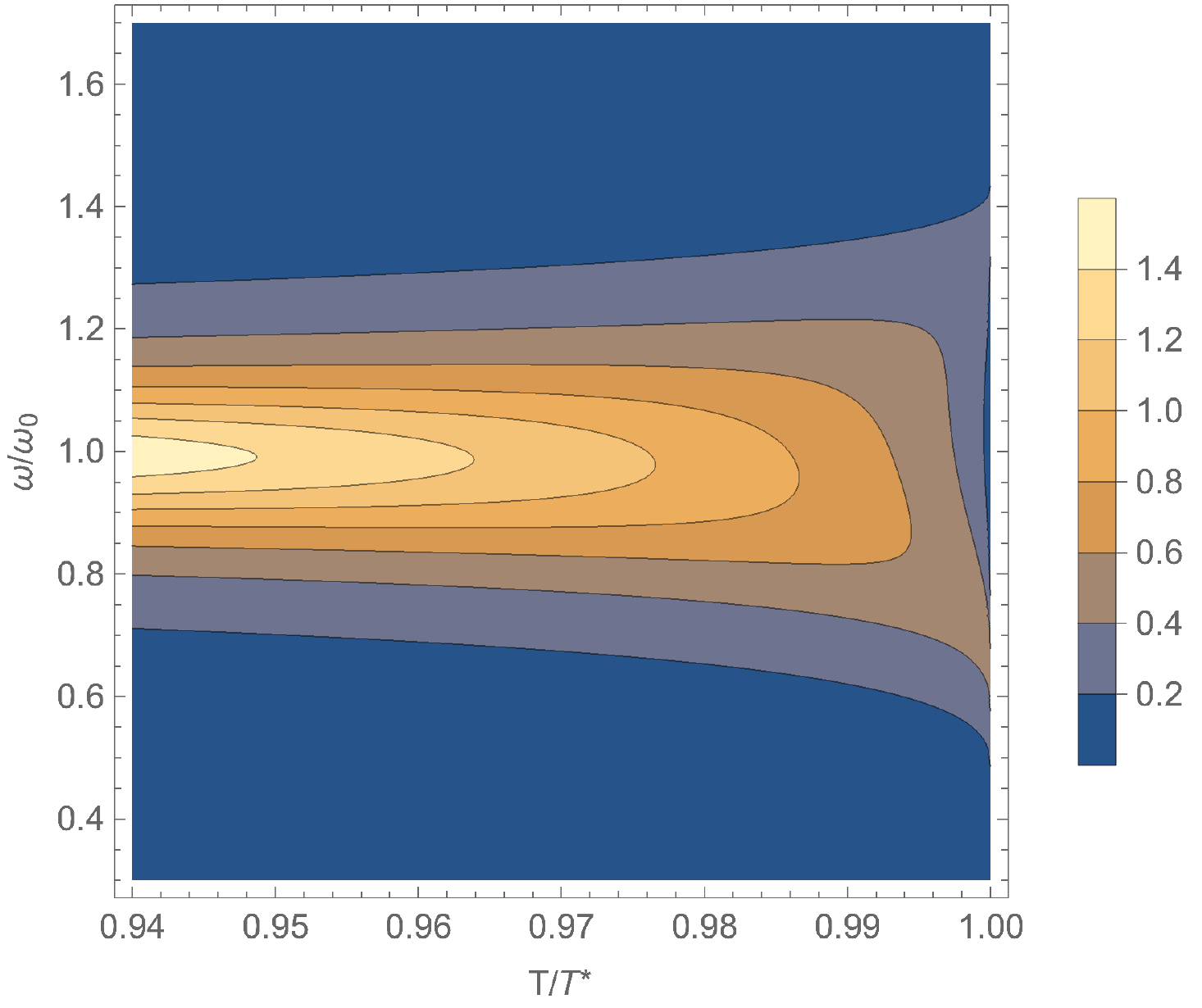}}
\caption{Q-ball scattered X-ray spectral line given by false colour intensity plots {\bf{a}}) of the Green's function $-\dfrac{1}{\pi}Im{D_R}(\omega, {\bf{p}})$ in Eq. (\ref{imag}) at different temperatures indicated by $T/T^*$ ratios as function of  frequency/wave-vector $p$ expressed as dimensionless frequency $\omega=cp$ in units of  incident X-ray frequency $\omega_0$, with dimensionless scattering amplitude $g=0.115$ defined as in Fig. \ref{21}; {\bf{b}}) is the same as a) but with higher resolution in temperature $T\approx T^*$. The picture is in good correspondence with experimental plot in Fig. 1 of the recently published X-ray diffraction data \cite{campi22} for the short-range dynamic CDW satellite peak intensity that appeared below $T^*$ temperature. \label{22}}
\end{figure}

The splitting of the peak into two at temperatures close to $T^*$ follows directly from Eq. (\ref{imag}) and will be evaluated below in analytic form.
The scattered X-rays line shape  is also possible to represent in a form of the false color plots in Fig. \ref{22} as generated from Eq. (\ref{imag}), with the splitting of the peaks in Fig. \ref{21} at temperatures very close to $T^*$ resulted in Fig. \ref{22}b. The figure Fig. \ref{22}a dramatically resembles experimental plots published recently \cite{campi22}.
Next, there are two opposite limits that can be used to derive analytically treatable consequences from the general Eq. (\ref{imag}). Namely, consider first the limit  $\pi\gamma M p\ll \kappa\approx R_{min}^{-1}$, where $R_{min}$ is the most probable Q-ball size close to $T^*$ temperature given by Eq. (\ref{RmT}). Then, Eq. (\ref{imag}) is reduced to a single Lorentzian:
  \begin{eqnarray}
&-\dfrac{1}{\pi}Im{D_R}(\omega, {\bf{p}})\approx 
\dfrac{\tilde{\Gamma}(\omega, {\bf{p}})}{ \dfrac{4}{p^2}\left(\dfrac{\omega}{c}-p\right)^2 +\tilde{\Gamma}^2(\omega, {\bf{p}})},\label{imaga}\\
&\tilde{\Gamma}(\omega, {\bf{p}})=\dfrac{2\pi^2\gamma_Q^2M^2{\Omega p}}{{c}\kappa^2},\label{GDptil}\\
&-\dfrac{1}{\pi}Im{D_R}(\omega, {\bf{p}})_{max}=\pi^{-1}\tilde{\Gamma}^{-1}(\omega, {\bf{p}}).\label{Amax}
\end{eqnarray}

\begin{figure}
\subfloat[a) ]{\includegraphics[width = 0.49\textwidth]{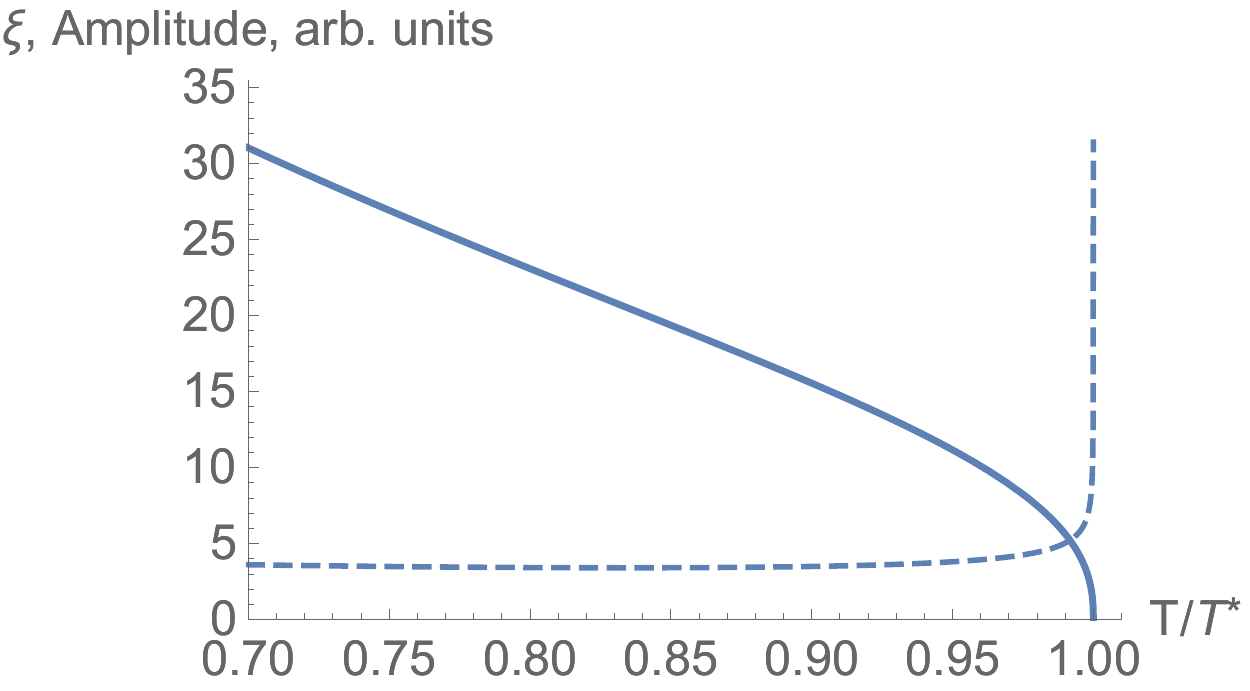}}
\subfloat[b) ]{\includegraphics[width = 0.49\textwidth]{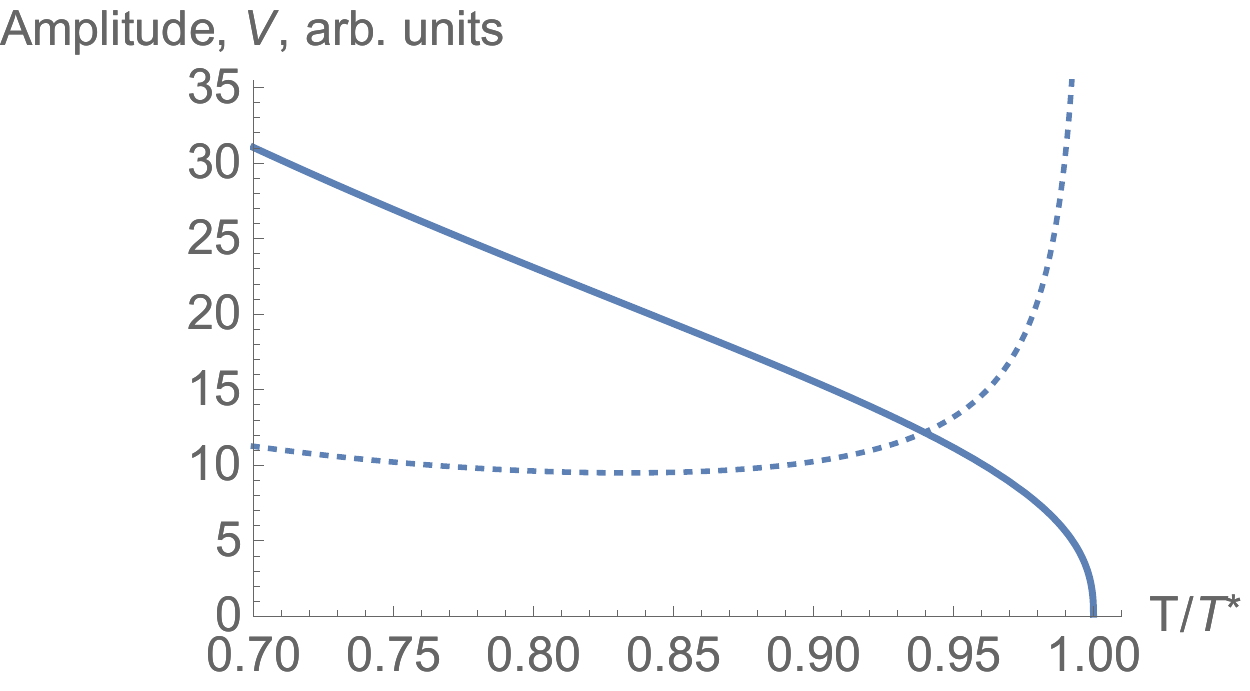}}
\caption{ a) Theoretical temperature dependence of the SDW/CDW single Q-ball scattered X-rays line peak amplitude and inverse of the peak width $\xi$:  
$-\dfrac{1}{\pi}Im{D_R}(\omega, {\bf{p}})_{max}$ (solid line),  $\xi=R_{min}$ (dashed line), as functions of reduced temperature expressed in units of  ${T/T^*}$; {\bf{b}}) solid line is the same as in a), dotted line is Q-ball volume $V_Q\propto R_{min}^3$ as function of reduced temperature ${T/T^*}$.  The pictures are in good correspondence with experimental plots in Fig. 2 of the recently published X-ray diffraction data \cite{campi22} for the short-range dynamic CDW X-rays scattering satellite peak intensity that appeared below $T^*$ temperature. \label{23}}
\end{figure}

\noindent Hence, the amplitude of the Lorentzian peak in Eq. (\ref{Amax}) equals $\pi^{-1}\tilde{\Gamma}^{-1}(\omega, {\bf{p}})$. Substituting 
temperature dependences of all the parameters entering the Lorentzian peak amplitude in Eq. (\ref{Amax}) from Eqs. (\ref{RmT}), (\ref{AmT}) one finds the following theoretical dependences plotted in Fig.\ref{23} of different characteristics of X-ray scattering.

\noindent These plots are in good qualitative correspondence with the experimental data of the X-ray scattering in high-$T_c$ cuprates HgBa$_2$CuO$_{4+y}$ \cite{campi22}.

\section{Q-ball mechanism of the $T$-linear temperature dependence of electrical resistivity in strange metal phase}
The above derivation of the X-ray diffraction by Q-balls could be now adjusted to show a cause of linear temperature dependence of  electrical resistivity of the high-T$_c$ superconductor in the Q-ball fluctuations phase described in this paper. 
\begin{figure}[H]
\includegraphics[width=0.45\linewidth]{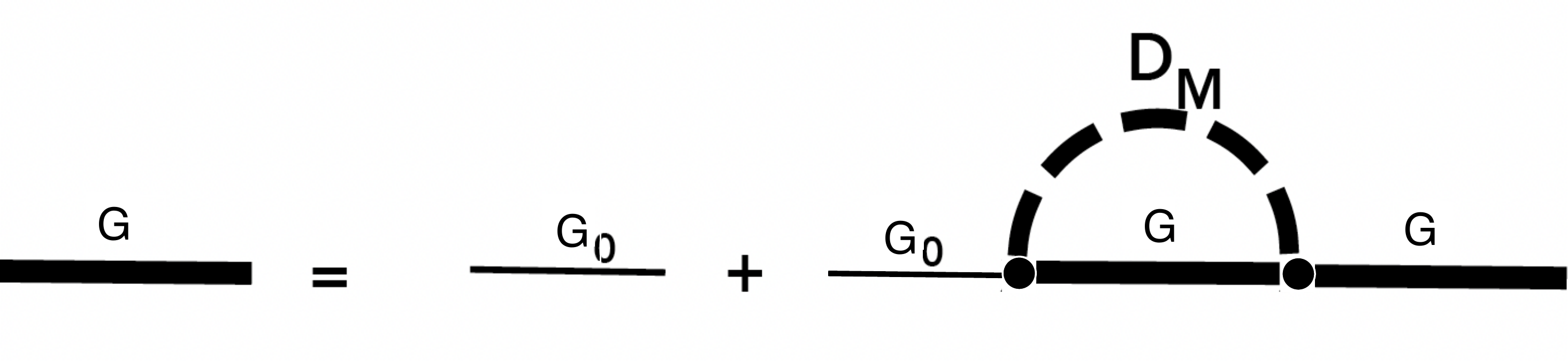}\hspace{0.2cm}
\caption{ The Dyson equation for a fermion scattering by Q-balls of CDW/SDW bosonic field : the dashed line is CDW/SDW Q-ball bosonic Euclidean field correlator $D_M^Q$ averaged over coordinates of Q-ball's centres in a crystal and Matsubara time zero-origin $\tau_0$. Heavy and thin lines are fermionic temperature Green's functions  $G({r}-{r}')$ and $G_0({r}-{r}')$ respectively. Dots are vertices of fermion- Q-ball field $M$ interaction Eq. (\ref{f}).} 
\label{Dys2}
\end{figure}

Consider now Fig. \ref{Dys2}, where now, contrary to Fig. \ref{Dys}, the heavy and thin lines are fermionic temperature Green's functions  $G({r}-{r}')$ and $G_0({r}-{r}')$ respectively, that depend on the differences of the $D+1$ coordinates after averaging over position of a Q-ball in space and Matsubara time origin $\tau_0$. Dots are vertices of fermion- Q-ball field $M$ interaction introduced in \cite{Mukhin(2022), Mukhin(2022_1)} via the expression:

\begin{eqnarray}
{S}_{f}=\int_0^{\beta}\int_Vd\tau d^D{\bf{r}}\sum_{{\bf{q}},\sigma}\left[ c^{+}_{{\bf{q}}\sigma}(\partial_\tau+\varepsilon_q)c_{{\bf{q}},\sigma}+\left(c^{+}_{{\bf{q+Q}_{DW}},\sigma}M(\tau,{\bf{r}})\sigma c_{{\bf{q}},\sigma}+H.c.\right)\right]\,. \label{f}
\end{eqnarray}

\noindent First, some rough evaluation is in order. Consider fermionic momentum uncertainty $\Delta p$ due to finite radius $R_Q$ of a Q-ball fluctuation and related uncertainty $\varepsilon_Q$ of fermionic excitation energy $\varepsilon$ in the vicinity of Fermi chemical potential: 
\begin{eqnarray}
&\Delta p\sim \dfrac{\hbar}{R_Q};\\
&\Delta\varepsilon(p)\equiv\varepsilon_Q\sim v_F\Delta p\sim v_F \dfrac{\hbar}{R_Q};\\
&\Delta t \Delta\varepsilon(p)\sim \tau_Q \Delta\varepsilon(p)\geq \hbar \Rightarrow \dfrac{1}{\tau_Q}\sim v_F \dfrac{1}{R_Q}\,,
\label{unc}
\end{eqnarray}
\noindent where $1/\tau_Q$ is fermionic quasiparticle decay rate due to Q-ball fluctuation. Now, using expression for the Q-ball conserved `Noether charge' Q Eq. \ref{Qs}, and temperature dependences for the M-field amplitude in Eq. (\ref{Mstar}) giving $M\propto T$ well below T$^*$ temperature, one finds T-linear dependence of $1/\tau_Q$:

\begin{eqnarray}
\dfrac{1}{\tau_Q}\sim v_F\dfrac{\hbar}{R_Q}\sim \left(\dfrac{\Omega M^24\pi}{3Q}\right)^{1/3}\propto \dfrac{T}{{N_Q}^{1/3}}\,,
\label{Tlin}
\end{eqnarray} 
\noindent where $N_Q=Q/g$ is temperature independent number of condensed charge/spin excitations forming a Q-ball. The temperature independence of $N_Q$ follows from the Boltzmann distribution $P(Q)$ of the "charges" Q via Q-balls  energies $E_{Q}$, given in  Eq. (\ref{aQ}), that prove to be linear temperature dependent: 

\begin{eqnarray}
P(Q)\sim \exp\left\{- \dfrac{E_Q}{k_BT}\right\}\sim\exp\left\{- \dfrac{2Q\Omega}{gk_BT}\right\}\equiv\exp\left\{- \dfrac{4\pi Q}{g}\right\},
\label{Qind}
\end{eqnarray}

\noindent where definition of bosonic Matsubara frequency $\Omega=2\pi Tn$, $n=1$ was used in accord with Eq. (\ref{SDWQ0}). Hence, using $1/\tau_Q$ in the Drude like kinetic equation for the fermionic quasiparticle momentum in external electric field  one finds T-linear electrical resistivity in the Q-ball fluctuations phase, in qualitative accord with high-T$_c$ cuprates behavior in the strange metal phase \cite{Zaanen}. Next, a more thorough derivation follows from the Dyson equation in Fig. \ref{Dys2} for the fermionic Green's function $G$ with M-field bosonic Green's function ${D^Q}_M$ given in Eq. (\ref{DDQ}):
\begin{eqnarray}
&&G(\omega,p)=\dfrac{1}{i\omega-\xi-\bar{G}};\quad \xi=v_F(|{p}|-p_F)\\
&&\bar{G}(\omega,p)=\sum_Qn_QM^2\dfrac{8\pi\kappa\hbar}{(2\pi)^3}\int d^3\vec{p}_1\left\{\dfrac{G(\vec{p}_1,\omega-\Omega)}{(\vec{p}-\vec{p}_1-\vec{Q})^2+\kappa^2}+\dfrac{G(\vec{p}_1,\omega+\Omega)}{(\vec{p}-\vec{p}_1+\vec{Q})^2+\kappa^2}\right\}=\nonumber\\
&&=\dfrac{4}{\pi}\sum_Qn_QM^2\varepsilon_Q\left(I_{+}+I_{-}\right)\,; \quad I_{\pm}=\int^{\infty}_{-\infty}\dfrac{d\xi}{\left[i(\omega\mp
\Omega)-\xi-\bar{G}_{\pm}\right]\left[{\varepsilon_Q}^2+(\xi-\xi_{\pm})^2\right]};\\
&&\xi_{\pm}=v_F(|\vec{p}\mp\vec{Q}|-p_F);\quad \bar{G}_{\pm}=\bar{G}(\omega,|\vec{p}\mp\vec{Q}|);\quad n_Q\propto\exp\left\{-\dfrac{4\pi Q}{g}\right\},
\label{DGn}
\end{eqnarray}
 \noindent that after analytic continuation to the real axis of $\omega$ gives retarded Green's function:
 \begin{eqnarray}
 &&G^{R}(\omega,p)=\dfrac{1}{i\omega-\xi-\bar{G}^{R}};\\
&&\bar{G}^{R}(\omega,p)={4}\sum_Qn_QM^2\varepsilon_Q\left\{\dfrac{1}{\omega+i\varepsilon_Q-\xi_{+}-{\bar{G}^{R}}_{+}}+\dfrac{1}{\omega+i\varepsilon_Q-\xi_{-}-{\bar{G}^{R}}_{-}}\right\}\,\\
&& {\bar{G}^{R}}_{\pm}=\bar{G}^{R}(\omega,|\vec{p}\mp\vec{Q}|).
\label{DGR}
\end{eqnarray}
\noindent The latter expression for $G^R$ finally leads to a scattering crossection renormalised expression for the quasi-particle life-time $\tau_Q$:
\begin{eqnarray}
\dfrac{1}{\tau_Q}\approx \dfrac{8\sum_Qn_QM^2}{\varepsilon_Q}\propto T\,,
 \label{DGtau}
\end{eqnarray} 
\noindent where the last T-linear estimate for $1/\tau_Q$ again follows from the mentioned above relations:$M\sim T$, $\varepsilon_Q\sim T$ and Eq. (\ref{DGn}) for $n_Q$.
\section{Conclusions}
\label{sec: fin}
To summarise, one concludes, that presented above theoretical results and their favourable comparison with experiment \cite{campi22,campi} indicate that X-ray diffraction makes "visible" the gas of Q-balls with Cooper pairs condensates below T*, and hence opens avenue for direct investigation of the thermodynamic quantum time crystals of CDW/SDW densities. In a particular picture related with high-T$_c$ scenario the vanishing density of superconducting condensates at T* leads to inflation of Q-balls sizes, that self-consistently suppresses X-ray Bragg's peak intensity close to Q-ball phase transition temperature. Linear temperature dependence of electrical resistivity in the Q-ball phase due to scattering of electrons on the condensed charge/spin fluctuations inside Q-balls is also demonstrated. The T-linear dependence of electrical resistivity arises due to inverse temperature dependence of the Q-ball radius as function of temperature in the strange metal phase.



\section{ACKNOWLEDGMENTS}
The author is grateful to prof. Antonio Bianconi for making available the experimental data on micro X-ray diffraction in high-T$_c$ cuprates prior to publication and to prof. Carlo Beenakker and his group for stimulating discussions of  the work. This research was in part supported by  Grant No. K2-2022-025 in the framework of the Increase Competitiveness Program of NUST MISIS.  




\end{document}